\documentclass[10pt, conference, letterpaper]{IEEEtran} 
\IEEEoverridecommandlockouts
\usepackage{cite}
\usepackage{amsmath,amssymb,amsfonts}
\usepackage{graphicx}
\usepackage{textcomp}
\usepackage{xcolor}
\usepackage{verbatim}
\usepackage{algorithm}
\usepackage{algpseudocode}
\usepackage{footnote}
\usepackage{CJKutf8}

\def\BibTeX{{\rm B\kern-.05em{\sc i\kern-.025em b}\kern-.08em
    T\kern-.1667em\lower.7ex\hbox{E}\kern-.125emX}}
\begin{document}
\author{\IEEEauthorblockN{Lei Yao\IEEEauthorrefmark{1}, Yin Xu\IEEEauthorrefmark{1}, Aimin Tang\IEEEauthorrefmark{1}, Taohe Chen\IEEEauthorrefmark{1}, Tianyao Ma\IEEEauthorrefmark{1}, Wenjun Zhang\IEEEauthorrefmark{1}}
\IEEEauthorblockA{\IEEEauthorrefmark{1} Shanghai Jiao Tong University,
Shanghai, China\\
Email: {\{yaoleiphd, xuyin, tangaiming, taohe.chen, mty0710, zhangwenjun\}@sjtu.edu.cn}
}
}
\title{Tensor Decomposition Based Mixed-Field Sensing for XL-MIMO AFDM Systems\thanks{This paper is supported in part by National Natural Science Foundation of China Program (62422111, 62371291, 62271316, 62431015), and in part by the Science and Technology Commission of Shanghai Municipality (Grant No. 25DP1500200).}}
\maketitle
\begin{abstract}
Integrated sensing and communications enabled by extremely large-scale MIMO (XL-MIMO) and affine frequency division multiplexing (AFDM) is a highly promising paradigm for vehicular networks. However, the near-field spherical wavefront distortions induce severe non-linear parameter coupling, while the highly dynamic scattering environments exacerbate mismatch errors. To address these critical challenges, this paper proposes a novel tensor-based sensing scheme for XL-MIMO AFDM systems. First, the received signals are reformulated into a tensor, followed by an efficient decomposition approach that exploits the inherent Vandermonde structure of the factor matrices. This allows parameters to be directly estimated from the decomposed matrices, effectively avoiding inter-parameter coupling. Subsequently, a symmetric decoupling and real-domain manifold optimization algorithm is proposed for angle of arrival estimation, circumventing the high-dimensional searches typically induced by near-field effects. Furthermore, a baseband reconstruction and analytical gradient-based algorithm is developed to perform delay-Doppler estimation in the continuous parameter domain, fundamentally eradicating the grid-mismatch errors inherent in high-mobility scenarios. With these decoupled factors, the remaining unknown angle of departure can be readily extracted. Extensive simulation results demonstrate that the proposed scheme achieves orders-of-magnitude improvements in delay-Doppler accuracy and eliminates the error floors in angular estimation that severely bottleneck state-of-the-art baselines.
\end{abstract}
\section{Introduction}
Vehicle-to-infrastructure (V2I) integrated sensing and communications (ISAC) is a core enabler for 6G intelligent transportation systems, supporting autonomous driving and smart traffic networks \cite{Du2023}\cite{Zhao2023}. By unifying communication and sensing on shared spectrum, it overcomes the limitations of traditional isolated systems.

However, V2I systems encounter challenges related to high mobility, which can severely degrade the sensing performance of conventional techniques. Severe Doppler spreads in orthogonal frequency division multiplexing (OFDM) destroy the orthogonality among subcarriers, which ultimately induces intercarrier interference (ICI). To address this, affine frequency division multiplexing (AFDM), based on the discrete affine Fourier transform (DAFT), has emerged as a superior alternative to existing waveforms \cite{Bemani2023}\cite{Chen2026}. AFDM utilizes two independently tunable discrete chirp parameters \cite{Yuan2025}, a flexibility that makes it particularly suitable for ISAC scenarios. On one hand, the tunable parameters enable AFDM to simultaneously support high-resolution delay-Doppler estimation for sensing and reliable one-dimensional equalization for communication. On the other hand, this unique algebraic mechanism allows AFDM to dynamically adapt to arbitrary delay-Doppler profiles and guarantee full diversity gain via simplified one-dimensional processing. Recent AFDM-ISAC studies have proposed parameter selection criteria for near-optimal ambiguity functions \cite{Zhu2024} and joint estimation schemes with low pilot overhead \cite{Ranasinghe2025}. However, these works have not yet explored the performance gains achievable through multi-antenna technologies.

In the spatial domain, equipping intelligent roadside units (RSUs) with extremely large-scale multiple-input multiple-output (XL-MIMO) arrays can unlock unprecedented spatial degrees of freedom, thereby substantially improving spectral efficiency and link reliability \cite{Cui2022}. However, XL-MIMO's large aperture extends the Rayleigh distance \cite{Zhu2025}, causing targets to span both near- and far-fields, forming a complex mixed-field topology.

So far, while extensive research has investigated ISAC within XL-MIMO architectures relying on conventional waveforms, the intersection of MIMO and AFDM for sensing remains largely unexplored. In \cite{Luo2026} and \cite{Yin2024}, MIMO-AFDM systems have been discussed; however, their focus remains exclusively on communication aspects, such as joint sparse graph receiver design in \cite{Luo2026} and diagonally reconstructed channel estimation in \cite{Yin2024}. Furthermore, while recent efforts have advanced AFDM-ISAC systems, they predominantly rely on far-field assumptions and small-scale setups, which fail to capture the severe parameter coupling induced by near-field spherical wavefronts in extremely large apertures. Although \cite{Luo2025} provides valuable insights into mixed-field sensing, there remains room to extend its applicable near-field range and further enhance the estimation robustness against potential grid mismatches. Therefore, it is urgent to design a high-precision and low-complexity sensing scheme for XL-MIMO AFDM systems that comprehensively takes near-field effects into account.

To address the aforementioned challenges, this paper proposes a unified tensor-based sensing framework tailored for XL-MIMO AFDM systems. First, the received signals are reformulated as a third-order tensor, and an efficient decomposition approach leveraging the inherent Vandermonde structure is developed to decouple the high-dimensional parameters. Subsequently, we propose a symmetric decoupling and real-domain manifold optimization (SD-RDMO) algorithm to estimate the angle of arrival (AoA), which effectively circumvents the high-dimensional search complexity induced by near-field effects. Furthermore, a baseband reconstruction and analytical exact gradient (BR-AEG) algorithm is developed for continuous-domain delay-Doppler estimation, fundamentally eliminating grid-mismatch errors in high-mobility scenarios. On this basis, angle of departure (AoD) is sequentially extracted. Simulation results verify that our scheme eliminates angular error floors and achieves orders-of-magnitude higher delay-Doppler accuracy than existing baselines.
\begin{figure}
    \centering
    \includegraphics[width=0.85\linewidth]{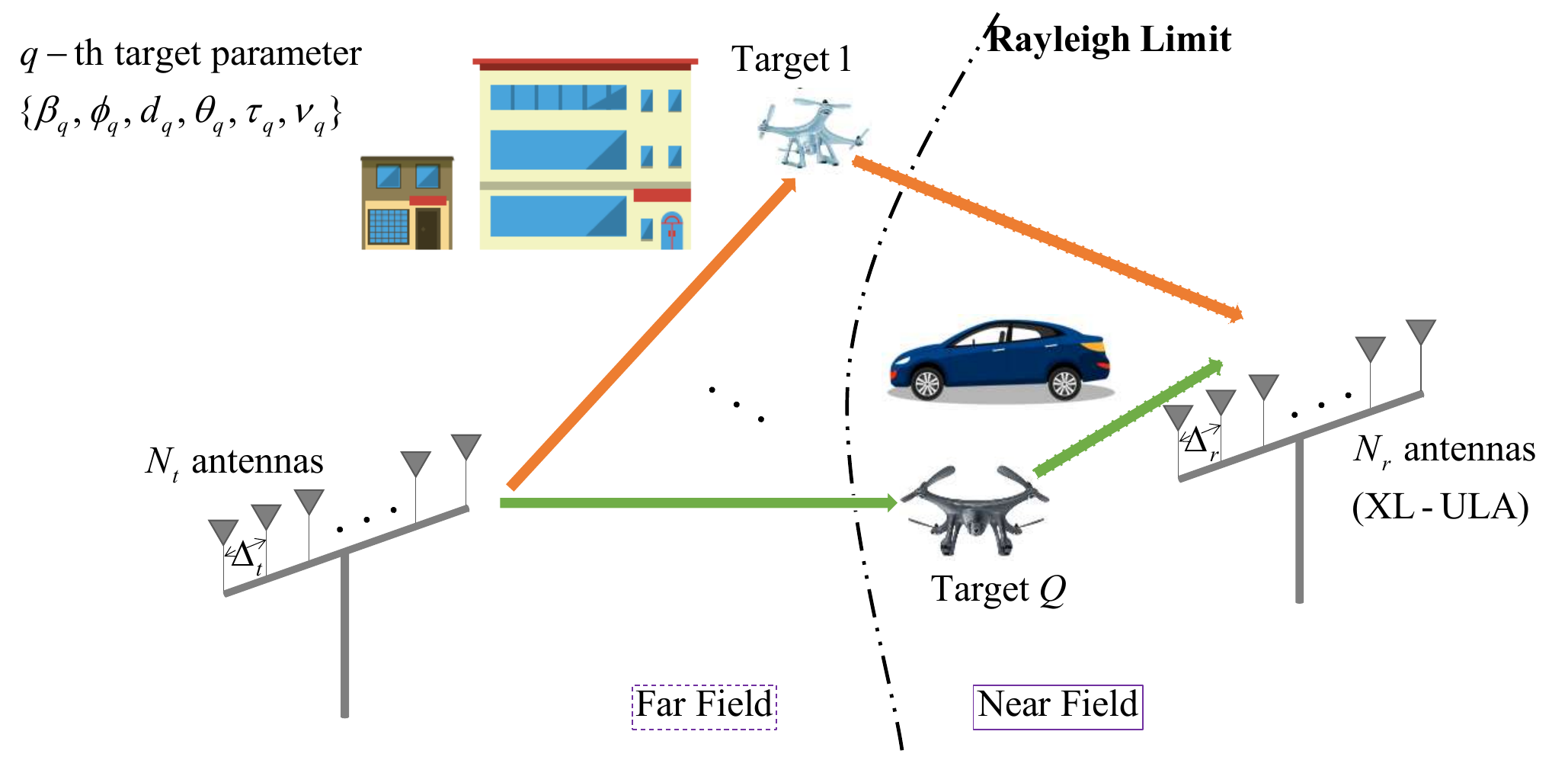}
    \caption{Bistatic Sensing System Model with XL-MIMO}
    \label{fig.SystemModel}
\end{figure}
\section{System Model}
This paper considers a bistatic AFDM sensing architecture for 6G intelligent transportation systems. As shown in Fig. \ref{fig.SystemModel}, there are $Q$ scattering targets with high dynamic characteristics in the mixed-field bistatic sensing region. The transmitter is equipped with an $N_t$-element uniform linear array (ULA). The receiver is an intelligent RSU deployed at a traffic hub. To achieve ultra high-resolution spatial sensing, the RSU is equipped with an extremely large-scale ULA (XL-ULA) of size $N_r$. Let $M$ be the number of subcarriers contained in a single symbol block, spanning a total bandwidth of $B$, which yields a subcarrier spacing of $\Delta f = B/M$. For the $N_t$ transmit antennas, they transmit the same Zadoff-Chu pilot sequence $\mathbf{x} \in \mathbb{C}^{M \times 1}$ in the DAFT domain. To enable the extraction of transmit spatial signatures and facilitate the subsequent tensor formulation, time-division multiplexing (TDM) is employed at the transmitter. 
\subsection{AFDM Modulation}

According to AFDM modulation, the mapping process of the time-domain transmit sequence $\mathbf{s} \in \mathbb{C}^{M \times 1}$ is defined as

\begin{equation}
\mathbf{s} = \mathbf{\Lambda}_{c_2}^H \mathbf{F}_M^H \mathbf{\Lambda}_{c_1}^H \mathbf{x},  
\end{equation}
where $\mathbf{F}_M \in \mathbb{C}^{M \times M}$ represents the normalized discrete Fourier transform matrix, $\mathbf{\Lambda}_{c_1} \triangleq \text{diag}(e^{-j2\pi c_1 m^2})_{m=0}^{M-1}$ and $\mathbf{\Lambda}_{c_2} \triangleq \text{diag}(e^{-j2\pi c_2 m^2})_{m=0}^{M-1}$ denote chirp phase rotation matrices. $c_1$ and $c_2$ represent the AFDM chirp parameters. To ensure the circular convolution property of the channel matrix in the discrete domain and to eliminate inter-carrier interference, a chirp periodic prefix (CPP) of length $L_\text{cpp}$ is appended to $\mathbf{s}$. Its $i$-th prefix sample is mapped as $\mathbf{s}[M+i]e^{-j2\pi c_1(M^2+2Mi)}$, where $i \in \{-L_\text{cpp}, \dots, -1\}$.

\subsection{Delay-Doppler Representation and AFDM Demodulation}
\label{subsection II.c}
Let $\tau_q$ and $\nu_q$ denote the echo delay and Doppler shift corresponding to target $q$. Accordingly, the normalized delay and normalized Doppler shift are defined as $\beta_q = \tau_q M \Delta f = l_q + \iota_q$ and $\alpha_q = \nu_q / \Delta f = k_q + \mu_q$, respectively. Here, $l_q \in (0,l_{\text{max}}]$ and $k_q\in [-k_{\text{max}},k_{\text{max}}]$ denote the integer parts, while $\iota_q, \mu_q \in [-0.5, 0.5)$ represent the corresponding fractional delay and fractional Doppler shifts. To prevent overlapping of distinct nonzero entries occupying the same position, $c_1$ is deliberately configured as $(2(k_\text{max}+k_v)+1)/2M$, where $k_v$ is the guard parameter according to the Doppler spread. $c_2$ is set to be an arbitrary irrational number or a rational number sufficiently small \cite{Bemani2023}.

At the receiver, the digital baseband processing first discards the redundant CPP of length $L_{\text{cpp}}$ and subsequently maps the discrete time-domain received signal back to the DAFT domain\cite{Bemani2023}. By incorporating the AFDM channel matrix structure, the manifold vector $\mathbf{v}_\text{DD}(\tau_q, \nu_q) \in \mathbb{C}^{M \times 1}$ for target $q$ can be analytically formulated as
\begin{equation}
\mathbf{v}_\text{DD}(\tau_q, \nu_q) = \mathbf{\Lambda}_{c_2} \mathbf{F}_M \mathbf{\Lambda}_{c_1} \mathbf{\Gamma}_{\text{CPP},q} \mathbf{\Delta}(\nu_q) \mathbf{\Pi}(\tau_q) \mathbf{s},    
\end{equation}
where we define $\mathbf{\Delta}(\nu_q) \triangleq \text{diag}(e^{-j2\pi m \alpha_q / M})_{m=0}^{M-1}$ to represent the discrete mapping matrix of the time-domain continuous Doppler rotation within the diagonal space. Similarly, $\mathbf{\Pi}(\tau_q) \triangleq \mathbf{F}_M^H \text{diag}(e^{-j2\pi m \beta_q / M})_{m=0}^{M-1} \mathbf{F}_M$ captures the linear phase shift effect induced by the signal delay. Furthermore, due to the different signal periodicity of the discrete chirp signals, $\mathbf{\Gamma}_{\text{CPP},q}$ is introduced to make the channel seemingly lie in a periodic domain for the integer delay component $l_q$. It is an $M \times M$ diagonal matrix explicitly defined as
\begin{equation}
\mathbf{\Gamma}_{\text{CPP},q} = \text{diag}\left( \begin{cases} e^{-j2\pi c_1(M^2-2M(l_q-m))} & m < l_q \\ 1 & m \ge l_q \end{cases} \right),
\label{eq.Gamma.CPP}
\end{equation}
where $m=0,\dots,M-1$. From \eqref{eq.Gamma.CPP}, we can see that whenever $2Mc_1$ is an integer and $M$ is even, $\mathbf{\Gamma}_{\text{CPP},q} = \mathbf{I}_M$.

\subsection{Manifolds Representation at Transmitter and Receiver}
Given that the aperture of the transmit antenna array is much smaller than the signal propagation distance, the transmitted signal is assumed to follow the far-field planar wavefront approximation. Letting the AoD of target $q$ be $\phi_q$, then the $n_t$-th component of the spatial steering vector $\mathbf{v}_\text{T}(\phi_q) \in \mathbb{C}^{N_t \times 1}$ for the transmit array is analytically expressed as

\begin{equation}
[\mathbf{v}_\text{T}(\phi_q)]_{n_t} = \exp\left( -j \frac{2\pi}{\lambda} n_t \Delta_t \sin\phi_q \right)  ,
\label{eq.farfield}
\end{equation}
where $\lambda$ and $\Delta_t$ denote the wavelength and the spacing between transmit antennas. 

Due to the extremely large physical aperture of the XL-ULA at the RSU, the classical Rayleigh distance is significantly extended.  Consequently, the $Q$ targets may scatter across both the near-field and far-field regions. To formulate a unified representation for this mixed-field scenario, we reconstruct the physical state space using a polar algebraic mapping manifold $\boldsymbol{\psi}_q \triangleq [\vartheta_q, \rho_q]^T$, so that the element of the mixed-field array manifold $\mathbf{v}_\text{R}(\boldsymbol{\psi}_q) \in \mathbb{C}^{N_r \times 1}$ at the RSU is defined as
\begin{equation}
[\mathbf{v}_\text{R}(\boldsymbol{\psi}_q)]_{n_r} = \exp\left( j \frac{2\pi}{\lambda} \left( n_r \Delta_r \vartheta_q - n_r^2 \Delta_r^2 \rho_q \right) \right),
\end{equation}
where $\Delta_r$ denotes the spacing between receiving antennas, $n_r \in \{-\frac{N_r-1}{2}, \dots, \frac{N_r-1}{2}\}$, indicating that e center antenna serves as the phase reference \cite{Cui2022}. The parameters $\vartheta_q$ and $\rho_q$ are explicitly determined by
\begin{equation}
\vartheta_q \triangleq \sin \theta_q, \quad \rho_q \triangleq \frac{1 - \vartheta_q^2}{2 d_q}.
\label{eq.AoD.mapping}
\end{equation}
Here, $\theta_q$ is the physical AoA of target $q$ relative to the RSU array center, $d_q$ is the distance between the $q$-th target and the reference antenna. This unified expression fundamentally originates from the exact spherical wavefront topology.
If target $q$ falls in the far-field region, the distance $d_q$ is sufficiently large such that $\rho_q$ approximates to zero. In this case, the manifold smoothly degenerates into the classical far-field planar wavefront steering vector just like \eqref{eq.farfield}.

Considering TDM framework, the $N_t$ transmit antennas emit the AFDM pilot sequence sequentially rather than concurrently. Each antenna $n_t \in \{1, 2, \dots, N_t\}$ occupies a dedicated time slot of duration $T_\text{sym}$. This temporal orthogonality guarantees that the echoes corresponding to different transmit antennas can be perfectly isolated at the ISAC receiver. Thus, the observation matrix $\mathbf{Y}_{n_t} \in \mathbb{C}^{N_r\times M}$ of the $n_{t}$ transmit antenna can be comprehensively modeled as
\begin{equation}
\mathbf{Y}_{n_t} = \sum_{q=1}^{Q} \gamma_q e^{j\Omega_{q,n_t}}[\mathbf{v}_\text{T}(\phi_q)]_{n_t}\mathbf{v}_\text{R}(\boldsymbol{\psi}_q) \mathbf{v}_{\text{DD}}^T(\tau_q, \nu_q)  + \mathbf{N}_{n_t},    
\end{equation}
where $\Omega_{q,n_t}=2\pi(n_t-1)\nu_qT_{\text{sym}}$ denotes the phase shift caused by the $q$-th target at the $n_t$-th slot. $\mathbf{N}_{n_t}\in \mathbb{C}^{N_r\times M}$ denotes the additive white Gaussian noise (AWGN) matrix and $\gamma_q \in \mathbb{C}$ denotes the complex scattering coefficient.

\section{Proposed Tensor-Based Parameters Estimation Algorithm}

\subsection{Tensor Reformulation and Decomposition}

To prevent the destruction of physical structural information and the coupling of multi-dimensional parameters caused by traditional matrix dimensionality reduction, we construct the observation model directly in a multi-dimensional algebraic space. By performing high-order stacking of the independent observation matrices $\{\mathbf{Y}_{n_t}\}_{n_t=1}^{N_t}$ \cite{Sorensen2013}, the global bistatic sensing observation tensor $\boldsymbol{\mathcal{Y}} \in \mathbb{C}^{N_r \times M \times N_t}$ can be constructed as
\begin{equation}
\boldsymbol{\mathcal{Y}} = \sum_{q=1}^{Q} \gamma_q  \mathbf{v}_\text{R}(\boldsymbol{\psi}_q) \circ \mathbf{v}_\text{DD}(\tau_q, \nu_q) \circ \tilde{\mathbf{v}}_\text{T}(\phi_q, \nu_q) + \boldsymbol{\mathcal{N}}, 
\label{eq.tensor.reformulation}
\end{equation}
where $[\tilde{\mathbf{v}}_t(\phi_q, \nu_q)]_{n_t}=e^{j\Omega_{q,n_t}}[\mathbf{v}_t(\phi_q)]_{n_t}$ denotes the coupling term between Doppler and AoD induced by TDM, $\circ$ represents the tensor outer product operator, and $\boldsymbol{\mathcal{N}} \in \mathbb{C}^{N_r \times M \times N_t}$ is the corresponding global noise tensor. To provide a more compact algebraic representation, we define the factor matrices associated with the three dimensions as
\begin{equation}
\begin{aligned}
\mathbf{A}_\text{R} &\triangleq [\mathbf{v}_\text{R}(\boldsymbol{\psi}_1), \dots, \mathbf{v}_\text{R}(\boldsymbol{\psi}_Q)]\in \mathbb{C}^{N_r \times Q} \\
\mathbf{A}_\text{DD} &\triangleq [\mathbf{v}_\text{DD}(\tau_1, \nu_1), \dots, \mathbf{v}_\text{DD}(\tau_Q, \nu_Q)]\in \mathbb{C}^{M \times Q} \\
\mathbf{A}_\text{T} &\triangleq [\tilde{\mathbf{v}}_\text{T}(\phi_1, \nu_1)), \dots, \tilde{\mathbf{v}}_\text{T}(\phi_Q, \nu_Q))]\in \mathbb{C}^{N_t \times Q}
\end{aligned}
.\label{eq.FactorMatrices}
\end{equation}
By further defining the scattering coefficient vector as $\boldsymbol{\gamma} \triangleq [\gamma_1, \dots, \gamma_Q]^T$, the observation tensor can be succinctly expressed using the Kruskal operator as $\boldsymbol{\mathcal{Y}} = \left[\!\left[ \boldsymbol{\gamma}; \mathbf{A}_\text{R}, \mathbf{A}_\text{DD}, \mathbf{A}_\text{T} \right]\!\right] + \boldsymbol{\mathcal{N}}$.

Define the full-dimensional physical parameter set to be estimated for the targets in the bistatic system as $\{ \theta_q, d_q, \phi_q, \tau_q, \nu_q, \gamma_q \}_{q=1}^Q$. The joint parameter estimation process is equivalent to minimize the global Frobenius norm reconstruction error, which is given by
\begin{equation}
    \text{minimize} \left\| \boldsymbol{\mathcal{Y}} - \sum_{q=1}^{Q} \gamma_q \mathbf{v}_\text{R}(\boldsymbol{\psi}_q) \circ \mathbf{v}_\text{DD}(\tau_q, \nu_q) \circ \tilde{\mathbf{v}}_\text{T}(\phi_q, \nu_q) \right\|_F^2.
\end{equation}

Thus, a potent strategy for parameter estimation involves leveraging the factor matrices, enabling the extraction of desired parameter information via canonical polyadic decomposition (CPD). A well-established method is the alternative least square (ALS) algorithm \cite{Yao2026}, which iteratively optimizes by updating one factor matrix at a time while keeping other parameters constant, albeit with high computational demands. Although the ALS-based approach provides an effective way to estimate the factor matrices, it fails to incorporate their structural constraints, thereby limiting the maximum number of resolvable targets. We propose to resolve the issue by fully leveraging the Vandermonde structure property of the factor matrices. 

To begin with, we obtain the mode-2 unfolding matrix of $\boldsymbol{\mathcal{Y}}$, which can be written as 
\begin{equation}
    \mathbf{Y}_{(2)}=\mathbf{A}_\text{DD} (\mathbf{A}_\text{T} \text{diag}(\boldsymbol{\gamma}) \odot \mathbf{A}_\text{R})^T + \mathbf{N}_{(2)},
\end{equation}
and $\mathbf{N}_{(2)}\in \mathbb{C}^{M \times N_t N_r}$ is the unfolding matrix of $\boldsymbol{\mathcal{N}}$ along its second dimension.
In \eqref{eq.FactorMatrices}, we observe that $\tilde{\mathbf{v}}_\text{T}(\phi_q, \nu_q)$ constitutes a component of a steering vector, preserving the power series configuration. Consequently, the factor matrix $\mathbf{A}_\text{T}$ demonstrates a Vandermonde structure with generators $\{e^{j(2\pi\nu_qT_{\text{sym}}-\frac{2\pi}{\lambda}\Delta_t\sin\phi_q)}\}_{q=1}^Q$. Building upon this insight, we incorporate the structural constraints inherent in factor matrices and introduce an algebraic-based estimation approach for these matrices. This method potentially circumvents the need for iterative processes and may even improve resolution when dealing with a larger number of targets  \cite{Yao2026}.
Specifically, we construct a cyclic selection matrix
$\mathbf{J}_{l_3} = \big[\mathbf{0}_{K_3 \times (l_3 - 1)}  \ \mathbf{I}_{K_3} \ \mathbf{0}_{K_3 \times (L_3 - l_3)} \big] \otimes \mathbf{I}_{N_r}  \in \mathbb{C}^{K_3 N_r  \times MN_r}$ and vary $l_3$ from 1 to $L_3$, yielding
\begin{equation}
	\begin{aligned}
		\mathbf{Y}_{S} & = \big[
		\mathbf{J}_{1} \mathbf{Y}_{(2)}^{T} \ \mathbf{J}_{2} \mathbf{Y}_{(2)}^{T},  \ldots ,
		\ \mathbf{J}_{L_3} \mathbf{Y}_{(2)}^{T}
		\big]  \\
		& = \big( \mathbf{A}_\text{T}^{(K_3)} \odot \mathbf{A}_\text{R} \big) \text{diag}(\boldsymbol{\gamma})
		\big( \mathbf{A}_\text{T}^{(L_3)} \odot \mathbf{A}_\text{DD} \big)^T + \mathbf{N}_{S},
	\end{aligned}
	\label{eq.cyclic.selection}
\end{equation}
where $K_3 + L_3 = N_t + 1$, and $\mathbf{A}_\text{T}^{(K)}$ denote the first $K$ rows of $\mathbf{A}_\text{T}$, and $\mathbf{N}_{S} \in \mathbb{C}^{K_3 N_r \times L_3M}$ is the corresponding noise term.

To estimate the factor matrices, we initiate the process with the computation of singular value decomposition (SVD). Specifically, we decompose $\mathbf{Y}_{S}$ as $\mathbf{Y}_{S} = \mathbf{U} \boldsymbol{\Sigma} \mathbf{V}^H = \mathbf{U}_{s} \boldsymbol{\Sigma}_{s} \mathbf{V}_{s}^H + \mathbf{U}_{n} \boldsymbol{\Sigma}_{n} \mathbf{V}_{n}^H$, where the $K$ principal singular vectors in $\mathbf{U}$ and $\mathbf{V}$ respectively span the signal subspaces $\mathbf{U}_{s}$ and $\mathbf{V}_{s}$, and $\boldsymbol{\Sigma}_{s}$ is a diagonal matrix composed of the $Q$ largest singular values. Define $\mathbf{U}_{1} = [\mathbf{U}_{s}]_{1:(K_3-1)N_r,:}$ and $\mathbf{U}_{2} = [\mathbf{U}_{s}]_{N_r + 1: K_3 N_r,:}$. We then compute the eigenvalue decomposition as $\mathbf{U}_{1}^{\dagger} \mathbf{U}_{2}= \mathbf{M} \mathbf{Z} \mathbf{M}^{-1}$. Consequently, the $q$-th column of the factor matrix ${\mathbf{A}}_\text{T}$ can be reconstructed as
\begin{equation}
    \hat{\mathbf{v}}_{\text{T},q} = \big[1, \hat{z}_{q}, \ldots, \hat{z}_{q}^{N_t-1}\big]^T,
\end{equation}
where $\hat{z}_{q} = [\mathbf{Z}]_{q,q}/\big| [\mathbf{Z}]_{q,q} \big|$ is the estimated generator. The other two factor matrices can be estimated, whose $q$-th columns can be obtained via
\begin{equation}
	\hat{\mathbf{v}}_{\text{R},q} = \big( \hat{\mathbf{v}}_{\text{T},q}^{(K_3) H} \otimes \mathbf{I}_{N_r} \big) \mathbf{U}_{s} \mathbf{m}_{q},
\end{equation}
\begin{equation}
    \hat{\mathbf{v}}_{\text{DD},q} =\big( \frac{ \hat{\mathbf{v}}_{\text{T},q}^{(L_3) H}  }{\hat{\mathbf{v}}_{\text{T},q}^{(L_3) H}\hat{\mathbf{v}}_{\text{T},q}^{(L_3) }}\otimes  \mathbf{I}_{M} \big)\mathbf{V}_{s}^{*} \boldsymbol{\Sigma}_{s} \mathbf{t}_{q}, \label{eq.a3}
\end{equation}
respectively, where $\mathbf{m}_{q}$ is the $q$-th column of $\mathbf{M}$ and 
$\mathbf{t}_{q}$ is the $q$-th column of $(\mathbf{M}^{-1})^{*}$.

\subsection{Sensing Parameters Extraction}
Up to now, our task has been transformed to extract the sensing parameters from aforementioned factor matrices, which necessitates addressing the three factor matrices.

\subsubsection{Mixed Field Parameters at the Receiver}
After obtaining the receive spatial manifold estimation vector $\hat{\mathbf{v}}_{\text{R},q} \in \mathbb{C}^{N_r \times 1}$ for the $q$-th target via tensor decomposition, the nonlinear coupling between the AoA and distance in the near-field spherical wavefront model incurs prohibitive computational complexity for direct two-dimensional spectral peak searching. To address this, we propose an estimation algorithm named symmetric decoupling and real-domain manifold optimization (SD-RDMO) to achieve high-precision parameter extraction through algebraic dimensionality reduction and continuous-domain optimization.

Exploiting the physical symmetry of the XL-ULA at the RSU, a spatial difference observation sequence is constructed. By extracting and conjugately multiplying the corresponding elements from the positive semi-axis $n_r \in \{1, \dots, \frac{N_r-1}{2}\}$ and the negative semi-axis $-n_r$ of the array, the decoupled observation value $h_{n_r}$ is obtained via
\begin{equation}
h_{n_r} = [\hat{\mathbf{v}}_{\text{R},q}]_{n_r} ([\hat{\mathbf{v}}_{\text{R},q}]_{-n_r})^* = \exp\left( j \frac{4\pi}{\lambda} n_r \Delta_r \vartheta_q \right).    
\end{equation}
This operation eliminates the quadratic term $\rho_q$ containing the distance information. It is worth noting that $\Delta_r$ should be set to $\lambda/4$ to avoid phase ambiguities. To suppress the impact of additive noise, a phase smoothing method is adopted to acquire the initial estimation of the angle mapping parameter $\vartheta_q^{(0)}$, which is expressed as
\begin{equation}
\vartheta_q^{(0)} = \frac{\lambda}{4\pi\Delta_r} \angle \left( \sum_{n_r=2}^{(N_r-1)/2} h_{n_r} h_{n_r-1}^* \right).    
\end{equation}

Subsequently, by performing quadratic phase compensation on the positive semi-axis signal, the phase observation value $w_{n_r}$ modulated solely by distance is extracted by
\begin{equation}
    w_{n_r} = \frac{[\hat{\mathbf{v}}_{\text{R},q}]_{n_r}}{[\hat{\mathbf{v}}_{\text{R},q}]_{0}} \cdot \exp\left(-j \frac{2\pi}{\lambda} n_r \Delta \vartheta_q^{(0)}\right)
\end{equation}
After extracting the continuous unwrapped phase of $w_{n_r}$, a weighted least squares fitting model is established to calculate the initial distance mapping parameter $\rho_q^{(0)}$:
\begin{equation}
\rho_q^{(0)} = \arg\min_{\rho_q} \sum_{n_r=1}^{(N_r-1)/2} \left( \angle w_{n_r} + \frac{2\pi}{\lambda} n_r^2 \Delta_r^2 \rho_q \right)^2.    
\end{equation}

To further improve parameter estimation accuracy and eliminate the cumulative error of phase unwrapping, a nonlinear optimization model is constructed with the state variable $\boldsymbol{\psi}_q = [\vartheta_q, \rho_q]^T$. We define the predicted signal at the $k$-th iteration as $\mathbf{m}_{\text{R}}^{(k)} = \delta_{1q}^{(k)} \mathbf{v}_\text{R}(\boldsymbol{\psi}_q^{(k)})$ and the complex residual vector as $\mathbf{e}^{(k)} = \hat{\mathbf{v}}_{\text{R},q} - \mathbf{m}_\text{R}^{(k)}$. $\delta_{1q}^{(k)}$ is the scale ambiguity factor caused by tensor decomposition, and is updated by $\delta_{1q}^{(k)} = \frac{(\mathbf{v}_\text{R}(\boldsymbol{\psi}_q^{(k)}))^H \hat{\mathbf{v}}_{\text{R},q}}{\|\mathbf{v}_\text{R}(\boldsymbol{\psi}_q^{(k)})\|_2^2}$. By deriving the partial derivatives of the predicted signal with respect to the physical parameters, the complex Jacobian matrix $\mathbf{J}_c^{(k)} \in \mathbb{C}^{N_r \times 4}$ is formulated as
\begin{equation}
  \mathbf{J}_c^{(k)} = \left[ \delta_{1q}^{(k)} \frac{\partial \mathbf{v}_\text{R}}{\partial \vartheta_q}, \quad \delta_{1q}^{(k)} \frac{\partial \mathbf{v}_\text{R}}{\partial \rho_q},\quad \mathbf{v}_\text{R},\quad j\mathbf{v}_\text{R} \right].
\end{equation}
where $\mathbf{v}_\text{R}$ is an abbreviation for $\mathbf{v}_\text{R}(\boldsymbol{\psi}_q^{(k)})$.

Due to the limitations of complex-domain differentiation, the complex observation space is equivalently projected into the real domain to construct the real-domain residual and Jacobian matrices as
\begin{equation}
\tilde{\mathbf{e}}^{(k)} = \begin{bmatrix} \Re(\mathbf{e}^{(k)}) \\ \Im(\mathbf{e}^{(k)}) \end{bmatrix}, \quad \tilde{\mathbf{J}}^{(k)} = \begin{bmatrix} \Re(\mathbf{J}_c^{(k)}) \\ \Im(\mathbf{J}_c^{(k)}) \end{bmatrix},
\label{eq.Re.Im.Part}
\end{equation}
where $\Re(\cdot)$ and $\Im(\cdot)$ denote the operations of taking the real and imaginary parts, respectively.
Based on this real-domain model, an adaptive damping term is introduced to solve for the iteration update step $\Delta \boldsymbol{\eta} \in \mathbb{R}^{4\times 1}$, given by
\begin{equation}
\Delta \boldsymbol{\eta} = \left( (\tilde{\mathbf{J}}^{(k)})^T \tilde{\mathbf{J}}^{(k)} + \mu \mathbf{I}_4 \right)^{-1} (\tilde{\mathbf{J}}^{(k)})^T \tilde{\mathbf{e}}^{(k)}.
\end{equation}
The state update $\boldsymbol{\psi}_q^{(k+1)} = \boldsymbol{\psi}_q^{(k)} + [\Delta \boldsymbol{\eta}[1], \Delta \boldsymbol{\eta}[2]]^T$ is executed until convergence. The high-precision physical parameters $\theta_q$ and $d_q$ are recovered through the inverse mapping \eqref{eq.AoD.mapping}.
\subsubsection{Delay-Doppler Estimation}
\label{subsubsection.delay-Doppler}
For the DAFT domain effective time-frequency manifold estimation vector $\hat{\mathbf{v}}_{\text{DD},q} \in \mathbb{C}^{M \times 1}$, traditional grid-search-based algorithms are prone to severe grid-mismatch errors in high-mobility scenarios. Therefore, we propose the baseband reconstruction and analytical exact gradient (BR-AEG) algorithm to achieve joint estimation in the continuous parameter domain.
To eliminate the discrete chirp phase coupling specific to the AFDM scheme, the observation vector $\hat{\mathbf{v}}_{\text{DD},q}$ is sequentially subjected to multiplication operations. This process equivalently restores the signal into the baseband time-domain observation sequence $\mathbf{y}_q$ as
$\mathbf{y}_q = \mathbf{\Lambda}_{c_1}^H \mathbf{F}_M^H \mathbf{\Lambda}_{c_2}^H \hat{\mathbf{v}}_{\text{DD},q}$
This operation avoids the boundary energy leakage caused by fractional delay and Doppler jumps. Defining the baseband prediction model as $\mathbf{v}(\beta, \alpha) = \mathbf{\Delta}(\alpha) \mathbf{\Pi}(\beta) \mathbf{s}$, a two-dimensional correlation search is performed over a low-resolution parameter grid set $(\mathcal{B}_{\text{grid}}, \mathcal{A}_{\text{grid}})$ to acquire the initial coordinates $[\beta_q^{(0)}, \alpha_q^{(0)}]^T$ according to
\begin{equation}
[\beta_q^{(0)}, \alpha_q^{(0)}]^T = \arg\max_{\beta \in \mathcal{B}_{\text{grid}}, \alpha \in \mathcal{A}_{\text{grid}}} \frac{\left| \mathbf{v}^H(\beta, \alpha) \mathbf{y}_q \right|^2}{| \mathbf{v }(\beta, \alpha) |^2}.
\end{equation}

To overcome the accuracy limitations of discrete grids, we then update the values of delay an Doppler through the procedures summarized in Algorithm \ref{Algorithm1}. 

\begin{algorithm}[htbp]
\caption{Delay and Doppler Updating in BR-AEG}
\begin{algorithmic}[1]
\Require $\mathbf{y}_q$, $\mathbf{s}$, $\beta_q^{(0)}, \alpha_q^{(0)}$, regularization parameter $\mu$, tolerance threshold $\epsilon$
\Ensure Estimated parameters $\hat{\beta}_q$ and $\hat{\alpha}_q$
\State Initialize iteration index $k = 0$
\State Initialize scale ambiguity factor 
\Statex $\delta_{2q}^{(0)} = \frac{(\mathbf{v}(\alpha_q^{(0)},\beta_q^{(0)}))^H \hat{\mathbf{v}}_{\text{DD}, q}}{\|(\mathbf{v}(\alpha_q^{(0)},\beta_q^{(0)}))\|_2^2}$
\State Construct initial joint state vector 
\Statex $\boldsymbol{\Theta}_q^{(0)} = [\beta_q^{(0)}, \alpha_q^{(0)}, \Re(\delta_{2q}^{(0)}), \Im(\delta_{2q}^{(0)})]^T$
\Repeat
    \State Compute the predicted signal 
    \Statex $\mathbf{m}_{\text{DD}}^{(k)} = \delta_{2q}^{(k)} \mathbf{v}(\beta_q^{(k)}, \alpha_q^{(k)})$
    \State Construct the complex Jacobian matrix 
    \Statex $\mathbf{J}_{\text{DD}}^{(k)} = [\frac{\partial \mathbf{m}_{\text{DD}}^{(k)}}{\partial \beta_q}, \frac{\partial \mathbf{m}_{\text{DD}}^{(k)}}{\partial \alpha_q}, \mathbf{v}, j\mathbf{v}]$ 
    \Statex and residual vector $\mathbf{e}_{\text{DD}}^{(k)} = \mathbf{y}_q - \mathbf{m}_{\text{DD}}^{(k)}$
    \State Compute $\tilde{\mathbf{J}}_{\text{DD}}^{(k)}$ and $\tilde{\mathbf{e}}_{\text{DD}}^{(k)}$ similar to \eqref{eq.Re.Im.Part}
    \State Calculate the update step:
    \Statex \qquad $\Delta \boldsymbol{\xi} = \left( (\tilde{\mathbf{J}}_{\text{DD}}^{(k)})^T \tilde{\mathbf{J}}_{\text{DD}}^{(k)} + \mu \mathbf{I}_4 \right)^{-1} (\tilde{\mathbf{J}}_{\text{DD}}^{(k)})^T \tilde{\mathbf{e}}_{\text{DD}}^{(k)}$
    \State Update the state variables $\boldsymbol{\Theta}_q^{(k+1)} = \boldsymbol{\Theta}_q^{(k)} + \Delta \boldsymbol{\xi}$
    \State $k = k + 1$
\Until{$\|\Delta \boldsymbol{\xi}\|_2 < \epsilon$}
\State \Return $\hat{\beta}_q = \boldsymbol{\Theta}_q^{(k)}[1]$ and $\hat{\alpha}_q = \boldsymbol{\Theta}_q^{(k)}[2]$
\end{algorithmic}
\label{Algorithm1}
\end{algorithm}
\subsubsection{AoD}
The AoD information can be extracted from $\hat{z}_q$ after the Doppler is estimated. Leveraging the bijective correspondence between variables $\hat{z}_q$ and $\hat{\phi}_q$, the parameter $\hat{\phi}_q$ can be accurately estimated by extracting the phase component from the complex variable $\hat{z}_q$, which is explicitly represented as
\begin{equation}
	{\hat{\phi}_q} =  \arcsin{(- \frac{\lambda }{{2\pi {\Delta_t}}}\angle {\hat {z}_q}+2\pi \hat{\alpha}_q\Delta f)}
	\label{v_estimate}.
\end{equation}

\subsection{Complexity Analysis}
In this subsection, the computational burden of the proposed tensor-based sensing scheme is provided. The calculation of the truncated SVD for $\mathbf{Y}_S$ requires $\mathcal{O}(Q K_3 L_3 N_r M)$ operations. Indeed, $\mathcal{O}(K_3 N_r Q^2)$ operations are needed to compute the pseudo-inverse of the signal subspace $\mathbf{U}_1$, while the eigenvalue decomposition procedure takes $\mathcal{O}(Q^3)$ operations. Furthermore, the SD-RDMO algorithm for AoA estimation takes $\mathcal{O}(Q N_r + Q I_{\mathrm{AoA}} N_r)$ operations, with $I_{\mathrm{AoA}}$ being the total number of iterations. Additionally, in the process of BR-AEG, the computational complexity connected with the baseband reconstruction and initial correlation search is $\mathcal{O}(Q M \log_2 M + Q G M)$, where $G$ denotes the total number of low-resolution grid points. Similarly, the computational complexity of the refinement in Algorithm \ref{Algorithm1} is $\mathcal{O}(Q I_{\mathrm{DD}} M)$, with $I_{\mathrm{DD}}$ denoting the associated number of iterations required for convergence.
\begin{figure}[t]
    \centering
    \begin{minipage}{0.48\columnwidth}
        \centering
        \includegraphics[width=\linewidth]{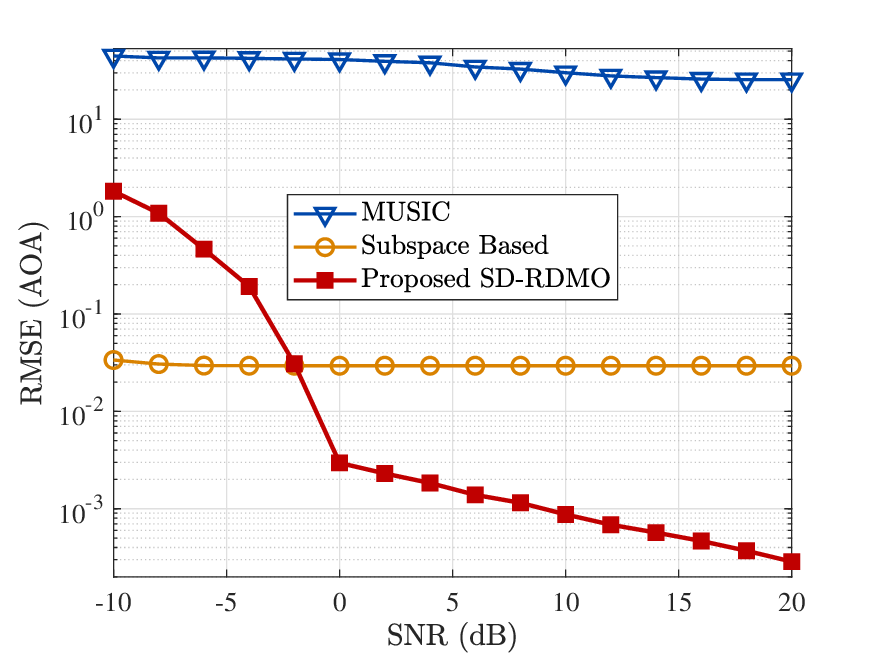}
        \caption{RMSE performance of AoA estimation versus SNR}
        \label{fig:RMSE-AoA}
    \end{minipage}\hfill
    \begin{minipage}{0.48\columnwidth}
        \centering
        \includegraphics[width=\linewidth]{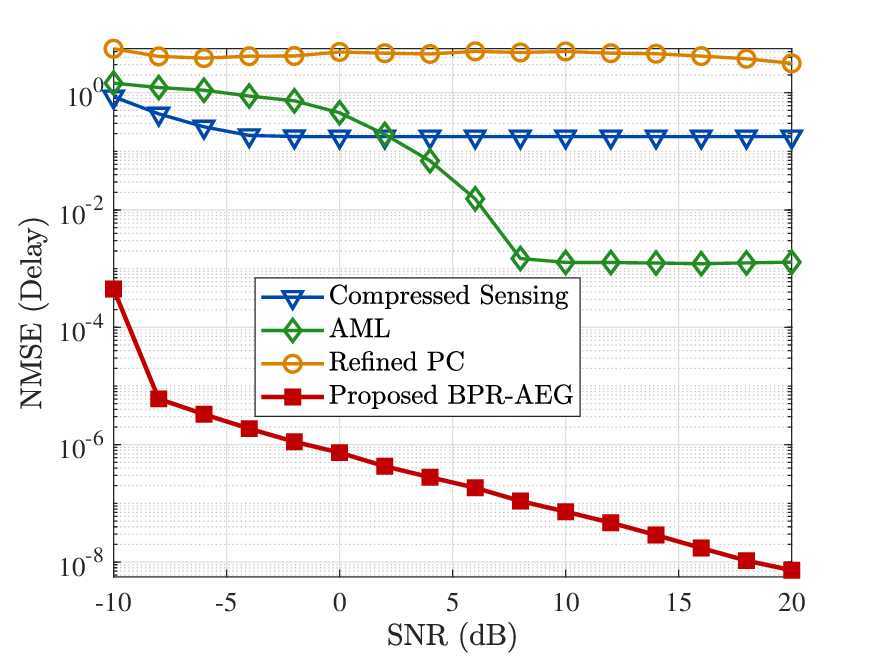}
        \caption{NMSE performance of delay estimation versus SNR}
        \label{fig:NMSE-delay}
    \end{minipage}
    
    \vspace{10pt} 
    
    \begin{minipage}{0.48\columnwidth}
        \centering
        \includegraphics[width=\linewidth]{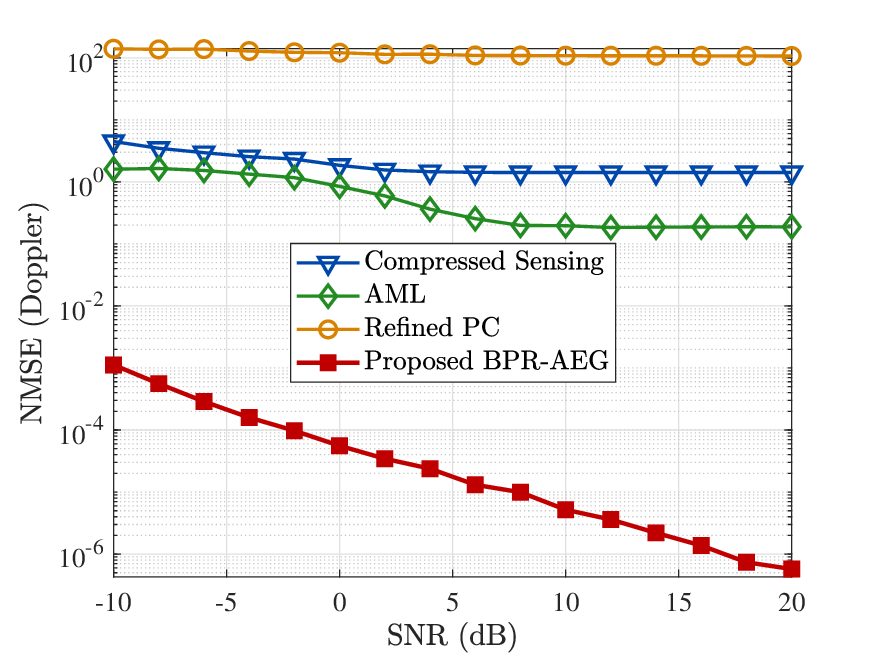}
        \caption{NMSE performance of doppler estimation versus SNR}
        \label{fig:NMSE-doppler}
    \end{minipage}\hfill
    \begin{minipage}{0.48\columnwidth}
        \centering
        \includegraphics[width=\linewidth]{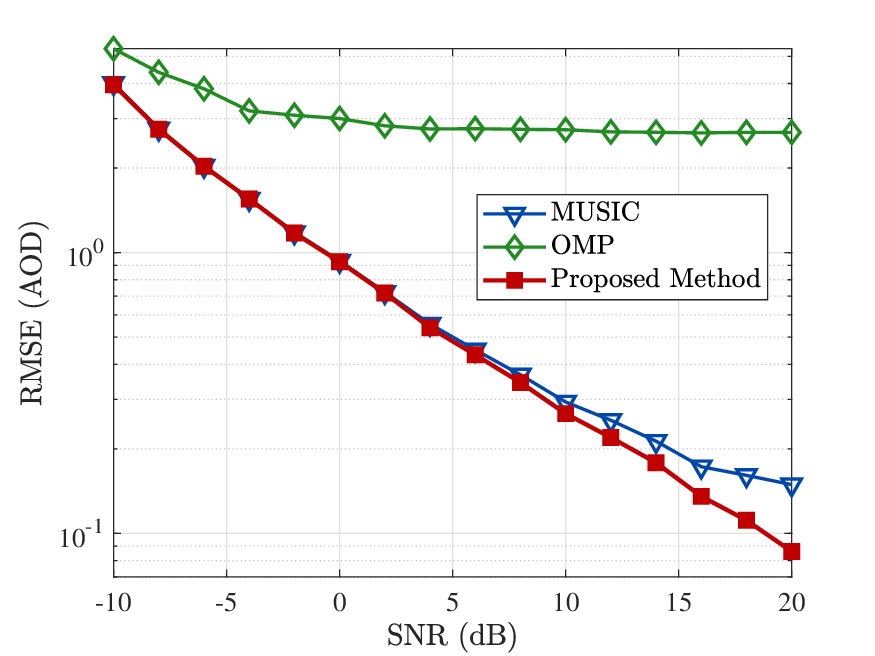}
        \caption{RMSE performance of AoD estimation versus SNR}
        \label{fig:RMSE-AoD}
    \end{minipage}
\end{figure}
\section{Simulation Results}
In this section, simulation results are presented to evaluate the performance of the proposed tensor-based sensing scheme. The carrier frequency of our V2I system is set to $60$ GHz, with a bandwidth of $B=10.24$ MHz, and $M=256$. For the transmitter, there are $N_t=8$ antennas, while there are $200$ antennas on each side of the reference antenna at the receiver, making the Rayleigh distance for the XL-ULA approximately $100$ meters. To strictly prevent spatial-Doppler ambiguity, the TDM time slot is compressed to $T_\text{sym}=28 \mu\text{s}$. Other parameters are set as $L_\text{cpp}=12$, $c_1=9/512$, and $c_2=1/10000$. In the simulation, three high-mobility targets are evaluated, comprising one near-field target and two far-field targets, at extremely high velocities of $60$, $-50$, and $80$ m/s, respectively. For the parameters AoD and AoA, the estimation performance is verified by root mean square error (RMSE). While for delay and doppler, it is more appropriate to use normalized mean square error (NMSE) for evaluation. For the parameter vector $\mathbf{h}$ to be estimated, its NMSE and RMSE are respectively defined as 
$\mathbb{E}\left[ \frac{\|\mathbf{h} - \hat{\mathbf{h}}\|_2^2}{\|\mathbf{h}\|_2^2} \right]$ and $
\sqrt{ \mathbb{E}\left[ \|\mathbf{h} - \hat{\mathbf{h}}\|_2^2 \right] },$ 
where $\hat{\mathbf{h}}$ denotes the corresponding estimated value.


As shown in Fig. \ref{fig:RMSE-AoA}, the RMSE of our proposed SD-RDMO AoA estimation scheme monotonically decreases with the escalation of signal-to-noise ratio (SNR). This asymptotic efficiency stems directly from our polar algebraic manifold formulation. The physical root of this failure lies in the rigid reliance of Multiple Signal Classification (MUSIC) on the planar wavefront assumption, resulting in smeared or entirely missing spatial spectral peaks. Furthermore, while the subspace based method \cite{Luo2025} partially addresses mixed-field sensing, its performance is fundamentally bottlenecked by the density of the partitioned spatial dictionary.


Figs. \ref{fig:NMSE-delay} and \ref{fig:NMSE-doppler} plot the NMSE of delay and Doppler estimation as functions of SNR. To rigorously evaluate our proposed scheme, we benchmark it against three representative baselines: the compressed sensing (CS) method \cite{Bemani2024-CS}, the alternating maximum likelihood (AML) algorithm \cite{Benzine2024-AML}, and the refined pulse compression (Refined PC) approach \cite{Luo2025}. In the considered high-mobility scenario, the grid-dependent CS method suffers from inevitable basis mismatch and energy leakage, while the Refined PC and AML approaches encounter severe resolution bottlenecks, both leading to prominent error floors at high SNR regimes. In stark contrast, the proposed BR-AEG scheme directly isolates the highly coupled time-frequency parameters by exploiting the intrinsic low-rank tensor structure. Consequently, it achieves a significantly lower NMSE and reduces the high-dimensional optimization to low-dimensional matrix operations, ensuring high performance with bounded complexity.

Fig. \ref{fig:RMSE-AoD} depicts the estimation performance of AoD versus SNR. In the scenario considered in this paper, the orthogonal matching pursuit (OMP) \cite{Lee2016-OMP} method experiences severe performance degradation. The fundamental bottleneck lies in its reliance on predefined discrete spatial-temporal dictionaries. Compared with the MUSIC \cite{Liu2020-MUSIC} method, the algorithm proposed in this paper achieves better performance. This is because the MUSIC process inherently annihilates the intrinsic multidimensional structural gains of the scattered electromagnetic waves.

\section{Conclusion}
In this paper, a tensor-based sensing framework tailored for XL-MIMO AFDM systems is developed. The received signals are reformulated as a third-order tensor, and an efficient decomposition approach exploiting the inherent Vandermonde structure is developed to decouple the high-dimensional parameters. To address the near-field effects, we introduce SD-RDMO algorithm for AoA estimation, which effectively circumvents the high-dimensional search complexity. For delay-Doppler estimation, BR-AEG algorithm is designed to operate in the continuous domain, fundamentally eliminating grid-mismatch errors in high-mobility scenarios. The remaining target parameter AoD is subsequently extracted in a sequential manner. Simulations confirm that the proposed framework achieves orders-of-magnitude lower delay-Doppler NMSE and eliminates the angular error floors common in existing grid-dependent methods.

\end{document}